\documentclass[a4paper,11pt]{article}
\pdfoutput=1 

\usepackage{jinstpub} 
\usepackage{booktabs} 
\usepackage{ dsfont }
\usepackage{mathtools}
\usepackage{calrsfs}
\DeclareMathAlphabet{\pazocal}{OMS}{zplm}{m}{n}
\usepackage{tikz}

\usepackage{subcaption}
\usepackage{upgreek}
\usepackage{array}
\usepackage{colortbl}
\usepackage{mwe}
\usepackage{tabularx}
\usepackage{calrsfs}
\DeclareMathAlphabet{\pazocal}{OMS}{zplm}{m}{n}
\usepackage{capt-of} 
\usepackage{comment}
\usepackage{xfrac}
\usepackage{abraces}
\usepackage{pict2e}

\title{Experimental measurement of the quality factor of a Fabry-Pérot open-cavity axion haloscope}

\author[a,b]{Juan F. Hernández-Cabrera,}
\author[c,a]{Javier De Miguel,}
\author[a,b]{E. Hernández-Suárez,}
\author[a,b]{Enrique Joven-Álvarez,}
\author[a,b]{H. Lorenzo-Hernández,}
\author[c]{Chiko Otani,}
\author[a]{Miguel A. Rapado-Tamarit,}
\author[a,b]{J. Alberto Rubiño-Martín,}
\author{on behalf of the DALI Collaboration}

\affiliation[a]{Instituto de Astrof\'isica de Canarias,\\E-38200 La Laguna, Tenerife, Spain}
\affiliation[b]{Departamento de Astrof\'isica, Universidad de La Laguna,\\E-38206 La Laguna, Tenerife, Spain}
\affiliation[c]{The Institute of Physical and Chemical Research (RIKEN),
Center for Advanced Photonics, 519-1399 Aramaki-Aoba, Aoba-ku, Sendai, Miyagi 980-0845, Japan}

\emailAdd{javier.miguelhernandez@riken.jp}

\abstract{The axion is a hypothetical boson arising from the most natural solution to the problem of charge and parity symmetry in the strong nuclear force. Moreover, this pseudoscalar emerges as a dark matter candidate in a parameter space extending several decades in mass. The Dark-photons \& Axion-Like particles Interferometer (DALI) is a proposal to search for axion dark matter in a range that remains under-examined. Currently in a design and prototyping phase, this haloscope is a multilayer Fabry-Pérot interferometer. A proof-of-principle experiment is performed to observe the resonance in a prototype. The test unveils a quality factor per open cavity of a few hundred over a bandwidth of the order of dozens of megahertz. The result elucidates a physics potential to find the, so far elusive, axion, in a sector which can simultaneously solve the symmetry problem in the strong interaction and the enigma of dark matter. }

\keywords{Dark Matter detectors, Performance of High Energy Physics Detectors}

\begin{document}
\maketitle
\flushbottom

\section{Introduction}
\label{sec:Introduction}
The quantum chromodynamics (QCD) axion is a pseudo-scalar Goldstone boson postulated by Weinberg and Wilczek 
\cite{PhysRevLett.40.223, PhysRevLett.40.279} as a consequence of the solution to
 the charge and parity symmetry conservation in the strong interaction suggested by Peccei and Quinn (PQ) \cite{PhysRevLett.38.1440}. Multiple extensions of the Standard Model (SM) of particle physics and string theory predict a set of particles similar to the axion, 
the so-called axion-like particles  (ALPs). Axions and ALPs mix with ordinary photons in a 
static, external, magnetic field. The axion has a light mass, induced by interactions with SM particles, 
which scales inversely to the PQ-field scale ($f_a$). For $f_a$ of the order of 10$^{12}$ GeV, or equivalently masses 
in the order of the $\upmu$eV, the axion is a well-grounded candidate for cold dark matter (DM) 
\cite{ABBOTT1983133, DINE1983137, 
PRESKILL1983127}. Since it emerged more than four decades ago, numerous searches for axion and ALPs have been carried out relying on theoretical-phenomenological hints, laboratory experiments and astronomy. A well-rounded summary of the different experimental approaches can be found in \cite{Baker:2013zta, Irastorza:2018dyq}; and a space where both ongoing projects and innovative proposals for new axion-seeking channels have a place is maintained at \cite{Ciaran}.  

The idea of employing a Fabry--Pérot resonator for scanning Galactic axions, as the concepts of resonant cavity haloscope—and helioscope—\textit{per se}, was suggested in \cite{Sikivie:1983ip}. Here, embedding grains or wires of a superconducting metal in a material transparent to microwave radiation provides the desired inhomogeneity of the magnetic field that enables the search for axion. Intended to perform a tunable broad band scan for axion at high-frequency, The Dark-photons \& Axion-Like particles Interferometer (DALI) \cite{DeMiguel:2020rpn, DeMiguel:2023nmz} incorporates a Fabry-P\'{e}rot (FP) resonator, composed of ceramic layers, instead of a resonant cavity, or rather than superconductig wire planes as adopted in Sikivie's approach. This set-up can extend the quest for the axion to a range of masses which remains poorly explored, in part since in the Sikivie's cavity-haloscope the resonant frequency, $\nu_0$, scales inversely on size, or the distance between the movable dielectric rods used for tuning, $d$, in the form $\nu_0\sim c/d$, $c$ being the speed of light; while the coherence volume, which boosts the output power, decreases accordingly. This makes it hardly practical to tune using electromechanical actuators above half a dozen gigahertz or, equivalently, a few dozen microelectronvolt of mass. In contrast, in FP haloscopes the resonance frequency is tuned by setting a wavelength fraction of distance between the ceramic layers, causing a constructive interference by reflection off a top mirror, the tuning results independent of the plate scale, provided its size is larger than the wavelength to avoid diffraction, rendering heavier axion scanning feasible. The output is spectrally modified compared to the input beam, allowing signal enhancement in relatively narrow frequency bands centered at a resonant frequency. The quality factor of a FP resonator, defined as 2$\pi$ times the ratio of the energy stored in the resonator and the energy loss per oscillation period, is $Q=\omega \tau_g$ \cite{Renk:2012}, where the group delay time, $\tau_g=-\mathrm{d}\phi/\mathrm{d}\omega$ with $\phi$ being the phase of the wavefront, is the average lifetime of a photon in the resonator. Equivalently, the group delay is associated with the decay time of the energy density of radiation in the interferometer, and adopts typical values of a few ns per layer. Other proposals for axion and dark photon \cite{Okun, Vilenkin} dark matter haloscopes based on the Fabry-Pérot resonator are MADMAX \cite{MADMAX:2019pub}, Orpheus \cite{Cervantes:2022epl} and LAMPOST \cite{Chiles:2021gxk}.  \\The operational principle of haloscopes is simple. Virialized axions of the Galactic halo can be converted into photons via inverse Primakoff effect \cite{Primakoff:1951iae} by action of a static magnetic field which contributes a virtual photon, $a+\gamma_{\mathrm{virt}} \rightarrow \gamma$, with a mass-frequency conversion factor $m_{a}/\upmu \mathrm{eV}\approx4.2 \nu_{\gamma}/\mathrm{GHz}$.  The power induced by ambient axions on the receiver of a Fabry-Pérot haloscope is

\begin{equation}
P \approx 10^5\,\mathrm{W} \times\frac{A}{\mathrm{m^2}}\times\left(\frac{B_0}{10\,\mathrm{T}}\right)^2\times Q \times g^2_{a\gamma}\times\left(\frac{\upmu \mathrm{eV}}{m_a}\right)^2\frac{\rho_a}{0.3\,\mathrm{GeVcm^{-3}}}\;,
\label{Eq.3}
\end{equation}
where $A$ is surface area of a plate, $B_0$ is magnetic flux density, $g_{a\gamma}$ is axion--photon coupling strength and, lastly, $\rho_{a}$ is the occupation of axion-like dark matter at the Earth, which saturates at several hundred megaelectronvolt per cubic centimetre. From Eq. \ref{Eq.3},  considering the weak coupling of QCD axion to standard photons, the signal generated by axion-to-photon conversion can be as low as a fraction of zeptowatt, which requires considerable strengthening to allow detection over an integration time containable on a human time scale. The  quality factor of a Fabry-Pérot device scales linearly with the number of consecutive layers. This law was already stated in  \cite{Millar}. Furthermore, it is straightforward if we consider the interferometer as a series of circuits where a capacitor adopts the form of two plates facing each other and an inductor is associated with the round trip of the photons inside each open-cavity \cite{2019JInst..14R8001D}. Interestingly, a higher relative permittivity, $\varepsilon_r$, results in a higher peak at the expense of a narrower full width at half maximum of the Lorentzian spectral feature originating from interferometry. A benchmark instantaneous scanning bandwidth of several dozens of megahertz, where the quality factor necessary to achieve QCD axion sensitivity, on the order of $Q\!\sim\!10^{4}$ , is tenable in practice is tenable according to simulations \cite{DeMiguel:2020rpn}, is targeted. The aim of this work is to verify experimentally the quality factor provided by a resonator made of multiple ceramic layers. The rest of the article is structured as follows. In Sec. \ref{sec:1}, the experimental approach is described. The experimental setup is reported in Sec. \ref{sec:2}. Results and a discussion are presented in Sec. \ref{sec:3} and Sec. \ref{sec:4}.

\section{The opto-mechanical tuner}
\label{sec:1}
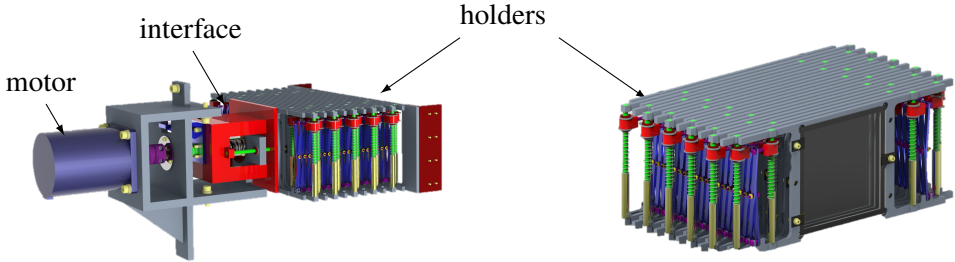
\begin{figure}[h]
    \centering
    \begin{minipage}[b]{0.45\linewidth}
        \centering
        \includegraphics [height=0.5\textwidth]{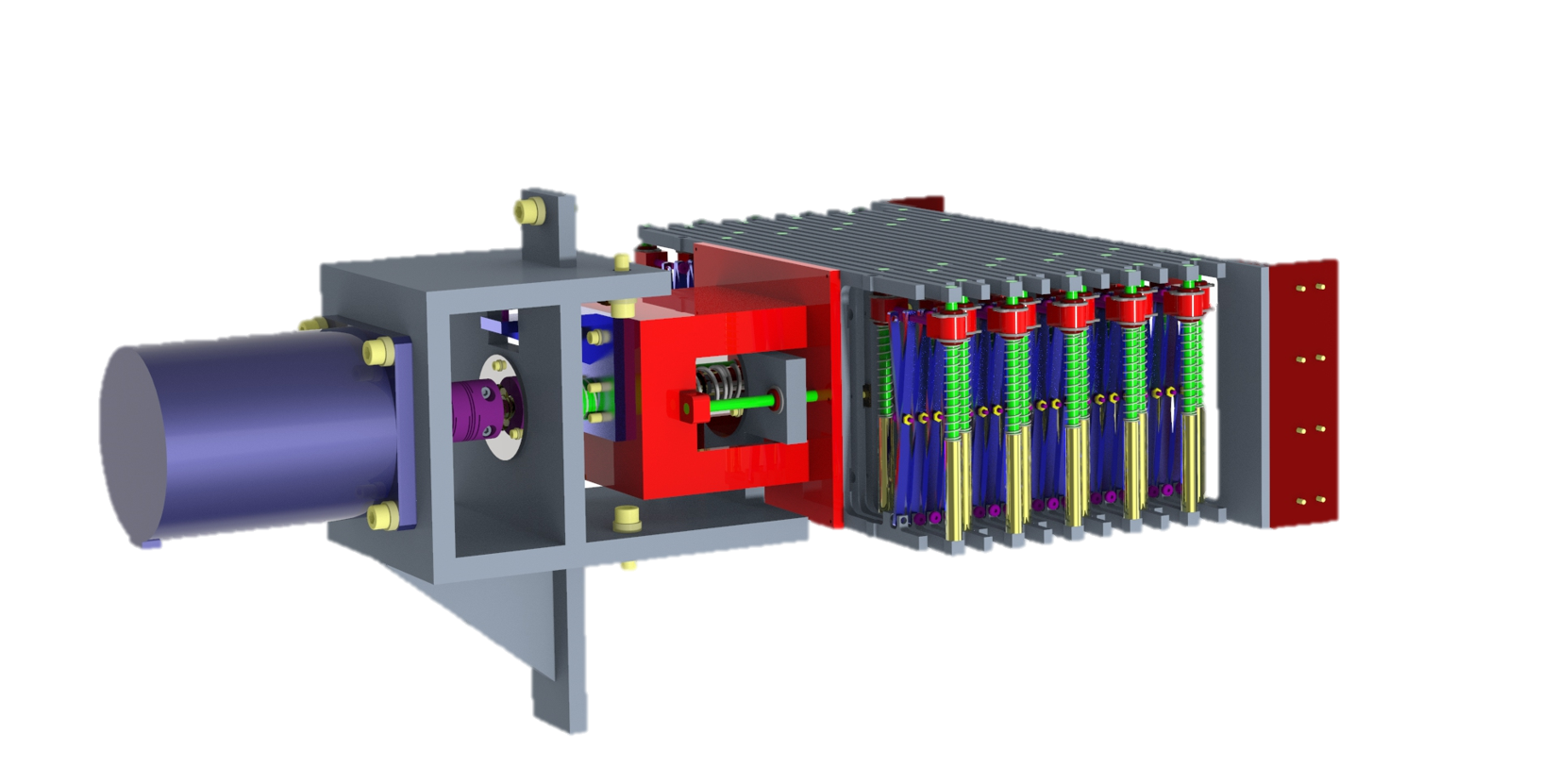}
    \put(-175,65){\vector(0.58,-1){7}}
     \put(-190,70){motor}
    \put(-120,85){\vector(0.58,-1){12}}
     \put(-140,90){interface}
         \put(17,90){\vector(0.999,-0.7){33}}
    \put(-20,90){\vector(-0.999,-0.7){30}}
     \put(-20,95){holders}

    \end{minipage}
    \begin{minipage}[b]{0.45\linewidth}
        \centering
        \includegraphics [height=0.5\textwidth]{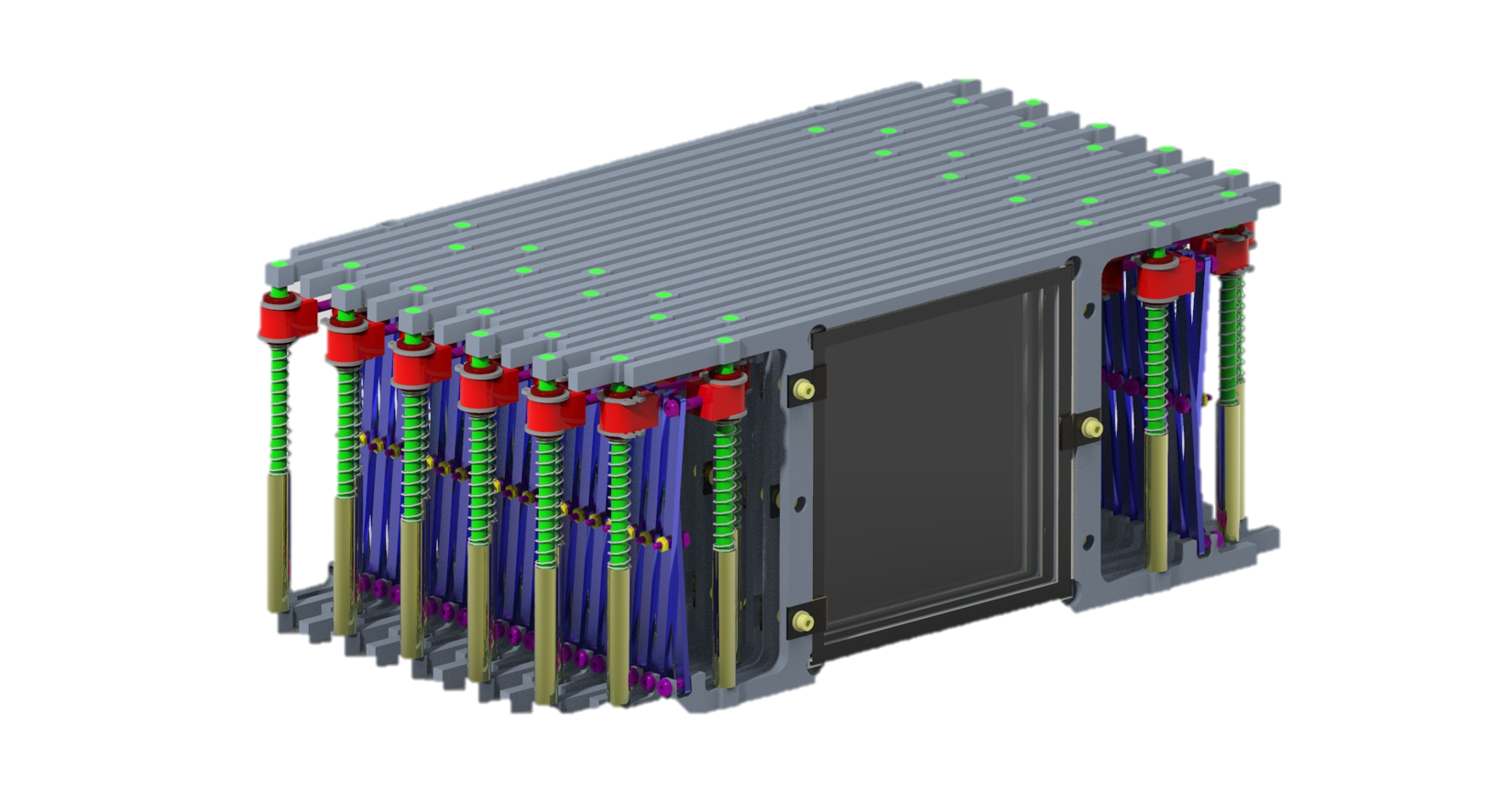}
    \end{minipage}
    \caption{\label{fig_1}Left: tunable Fabry--Pérot resonator. Motor displaces the interface. Plates move in unison by means of a scissor plus a bearing attached to each holder. Preloading with springs on each shaft equalizes the position uncertainty. Right: scissor tuning mechanism and holders, detail.}
\end{figure}

A conceptual design of a Fabry-Pérot resonator for DALI is shown in Fig. \ref{fig_1}. The `scissor' concept is used to allow adjacent plates to move simultaneously while maintaining the same spacing, which allows for tuning and, at the same time, restricts unwanted degrees of freedom. Three-dimensional finite element method (FEM) simulations\footnote{CST Studio Suite v. 2021.00 licensed by IAC.} powered with adaptive mesh-refinement have been used to establish the mechanical error budget, or uncertainty in the layer positioning, at some $\epsilon\lesssim\lambda/500$, $\lambda$ being the scanning wavelength. This error budget does not represent a particularly pressing challenge, specially below a hundred microelectronvolt of mass or, equivalently, several dozen gigahertz frequency. All devices used are manufactured from non-magnetic materials, such as aluminium, copper or non-magnetic austenitic steel, and thus magnetic forces on the actuators and mechanisms vanish allowing for an optimal operational efficiency. We have divided the opto-mechanical system in Fig. \ref{fig_1} into two components, namely, optics and mechanics. The study of optics is the scope of this manuscript. The device under test is composed of a fixed-plate FP resonator which allows us to measure the resonance around a resonant frequency of about three tens of gigahertz, with $N$ = 20 plates made of yttria-stabilized tetragonal zirconium oxide—Zr$\mathrm{O}_2$ \cite{Molla}, fiducial electric permittivity  $\varepsilon_r\sim20-40$; losses $\mathord{\mathrm{tan}}\,\delta\sim5\times10^{-3}$; surface roughness $R_{\mathrm{a}}<0.1$  $\upmu$m—of dimensions 10 cm $\times$ 10 cm area and 1 mm thickness. Two grooved aluminium layer holders have been manufactured to fix the plates in position with spacings of 6.04 mm and 6.21 mm, with a manufacturing error lower than a few dozen micrometers, in order to resonate at frequencies $\sim$500 MHz apart. 

\section{Experiment set-up}
\label{sec:2}

\begin{figure}[h]
    \begin{center}
    \resizebox{.9\textwidth}{!}{%
    \includegraphics[height=2cm]{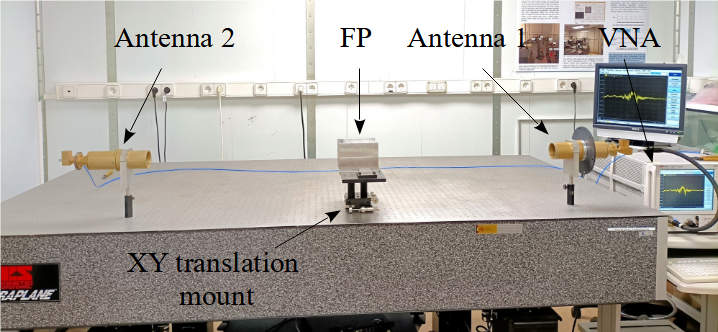}%
    \quad
    \includegraphics[height=2cm]{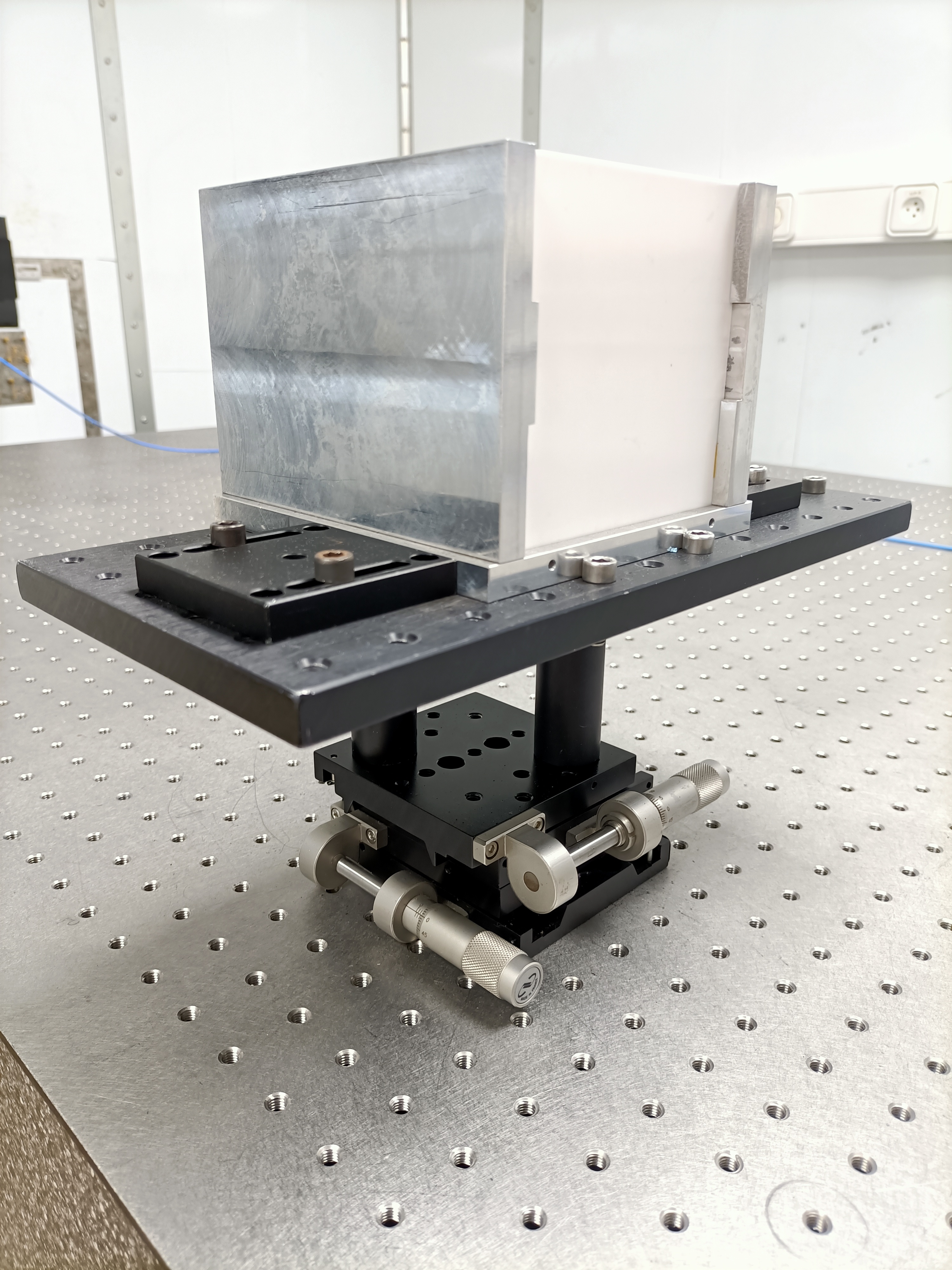}%
    \quad
    \includegraphics[height=2cm]{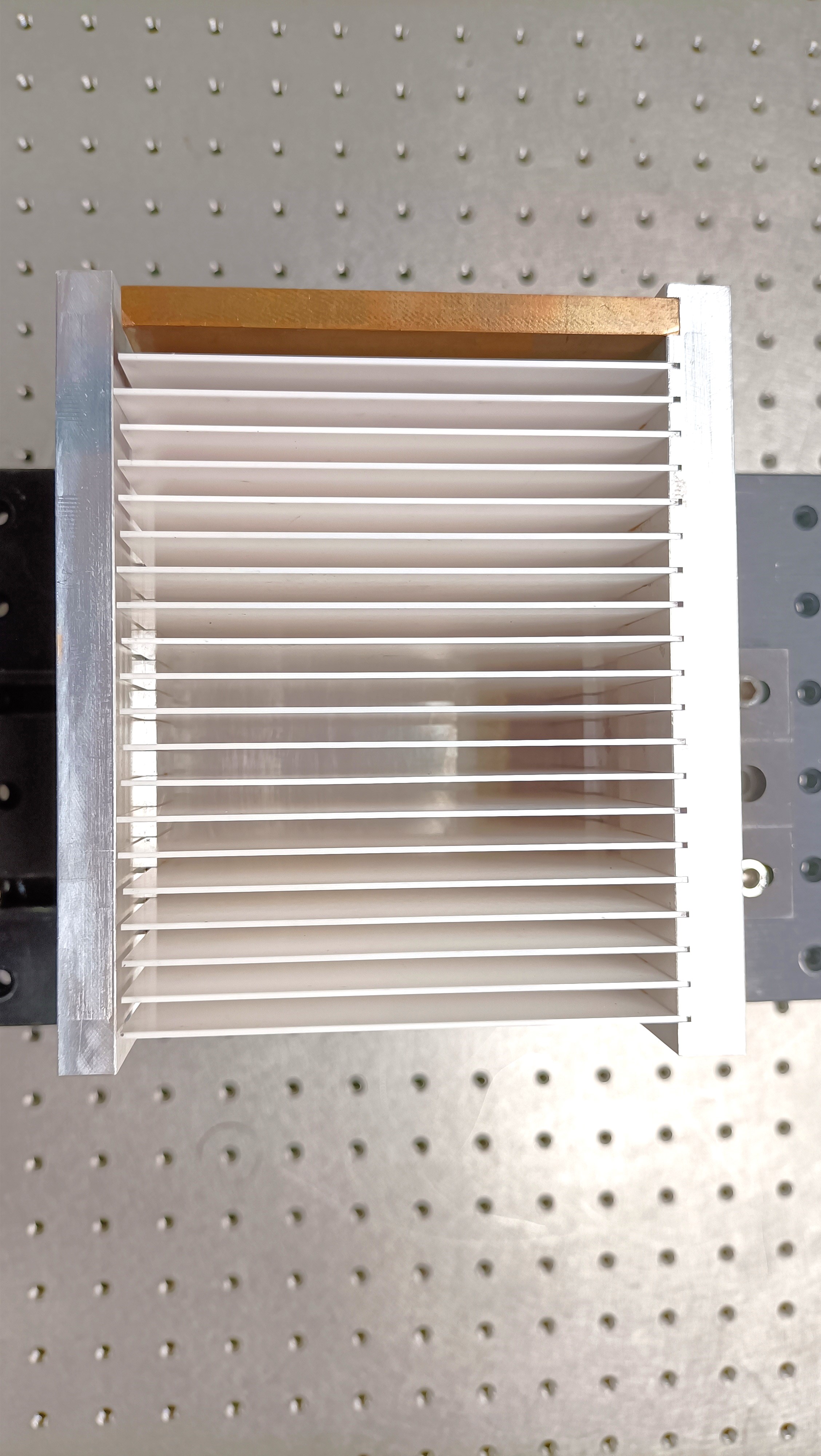}%
    }
    \end{center}
    \caption{Left: A photograph of the experimental setup. Center: Detail of the XY translation mount used to restrict the orientation of the FP resonator as well as to adjust its position. Right: Top view of the FP resonator including the top mirror. }
    \label{fig_2}
\end{figure}

We performed an experiment to obtain a measurement of amplitude and group delay over frequency across the resonator with a free-space optical and electronic setup that can provide adequate repeatability across measurements. An optical table located in a Faraday-cage shielded room has been used to fasten two antennas—cylindrical corrugated horns of a high directionality and gain \cite{QUIJOTE:2015npn}—pointing towards the resonator, which has been installed on an XY translation mount adjusted in such a way that both waveguide axes and the line defined by all plate geometrical centers are coincident lines located 200 mm above the optical table surface with an error of $\lesssim$1 mm (see Fig. \ref{fig_2}). Scattering parameters have been measured with a Keysight N5245A PNA-X vector network analyzer (VNA) driving signals through the resonator from the antennas. 
The waveguide outer rims have been separated far enough from the outer plates to ensure they are in the Fresnel region, i.e., $0.62 \sqrt{D^3 / \lambda}\sim$86 mm at the highest frequency; $D$ being the antenna aperture. A complete restriction of all degrees of freedom has been achieved in the following manner. The resonator orientation and positioning and the antennas positioning are restricted by the optical setup; alignment of the waveguide axes was ensured by verifying parallelism between the antenna outer rims and the optical table grid with an error of $\lesssim$$1^{\circ}$. The zirconia plates are restricted in their position and centered in the grooves by using polyimide film with silicone adhesive in their contact points with the holder. 

Broadband scattering parameter measurements with a $\sim$2.5 MHz resolution have been obtained in a data acquisition campaign by varying the number of installed zirconia layers from 0 to 20. After calibrating the VNA at the antenna connectors with a short-open-load-through procedure, raw data with and without the top mirror have been recorded for each layer count. This procedure was repeated for both fixed holders. The two-layer transmission measurement without the top mirror is a Fabry-Pérot resonator whose transmissivity can be characterized by dividing the $S_{21}$ data by the $S_{21}$ measured with an empty layer resonator—zero layers and no top mirror—in order to eliminate the attenuation originated in the transmission medium and subtract any phase delay. The resulting structure is identified to verify the presence, location and characteristics of the resonant feature. This procedure is repeated for all remaining layer counts up to twenty. The zero-layer reflectivity measurement with the top mirror is used to calculate the delay of the reflection from the antenna to the mirror and back $\tau_{g}^{(0)}$ as the average of the $S_{11}$ group delay in the interest frequency band, which will be subtracted from all calculated group delay structures to eliminate this offset. It has been deemed unnecessary to divide the reflectivity data by a background measurement-such as the zero-layer $S_{11}$ data with the top mirror to avoid combining errors from both measurements or potentially inducing artifacts in the numerical computation of the unwrapped phase derivative to calculate the group delay. It is not required to eliminate the attenuation originated in the transmission medium, since this does not have a bearing on the group delay, and the phase delay introduced by the medium can be cancelled by subtracting $\tau_{g}^{(0)}$ as previously described. Gate limits are modified to isolate the resonant structure at its location away from at $\sim$ 6 ns delay. On the contrary, cancelling medium attenuation frequency-wise is necessary for an accurate measure of the peak transmissivity of the two-layer FP resonator, and consequently a division has been performed in this case. Any small features in the resulting transmissivity pattern are eliminated by means of time-domain gating, and artifacts introduced in the phase data are not of concern since only the magnitude of the transmissivity is of interest.

Reflectivity measurements with $N = 1, \ldots, 20$ plates are gated in the time domain with a Tukey window \cite{Prabhu:2018} whose lower bound captures the beginning of the reflection. The higher bound is swept in a 9.5 ns range to identify an area in which the last peak group delay is stable. The average value in a range spanning 2 ns to each side of the minimum value is selected for analysis, avoiding peak-like structures. The quality factor is calculated as the product of the peak group delay and the resonant angular frequency.

\section{Results}
\label{sec:3}
\begin{figure}[h]
    \centering
    \includegraphics[width=0.85\textwidth]{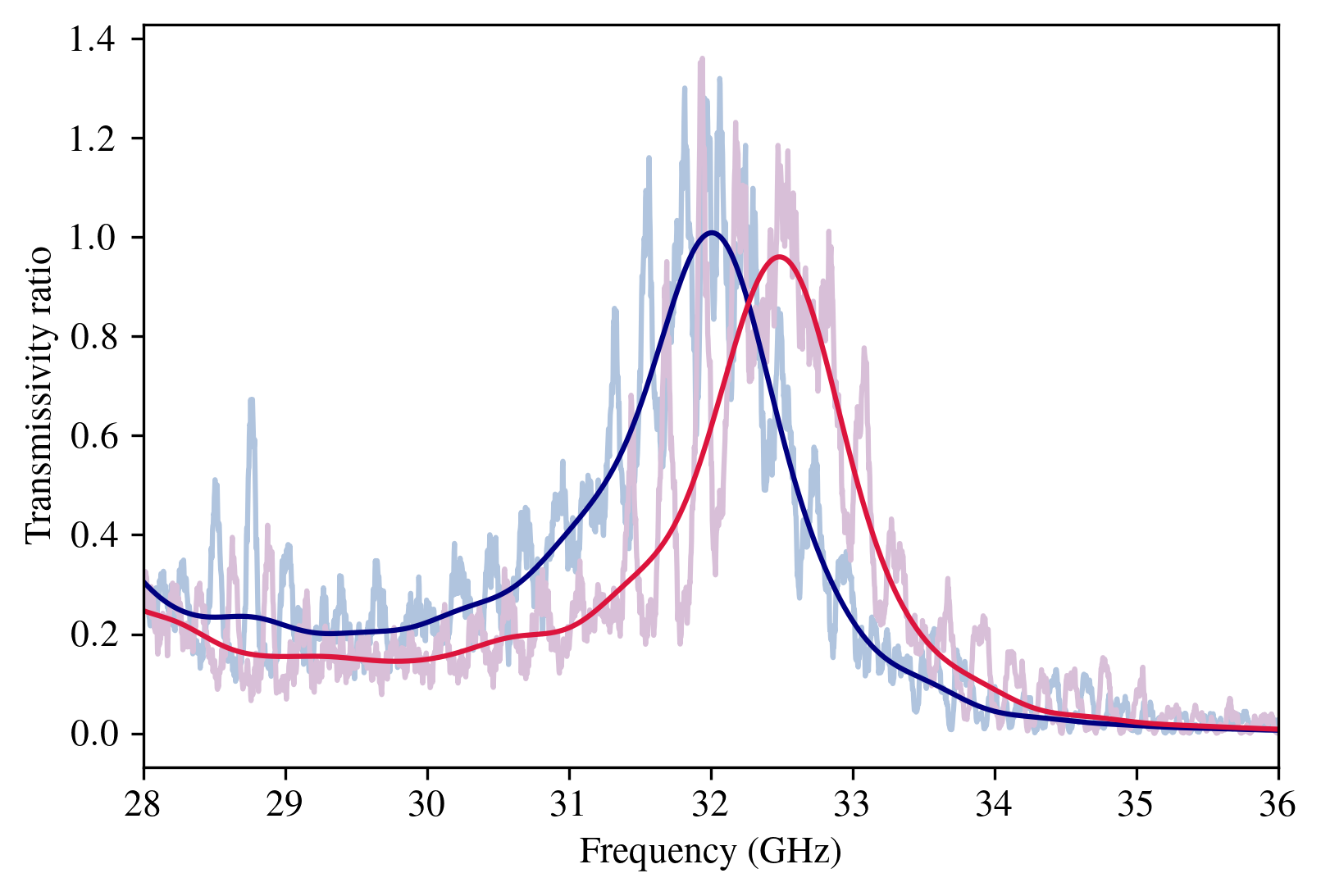}
    \includegraphics[width=0.85\textwidth]{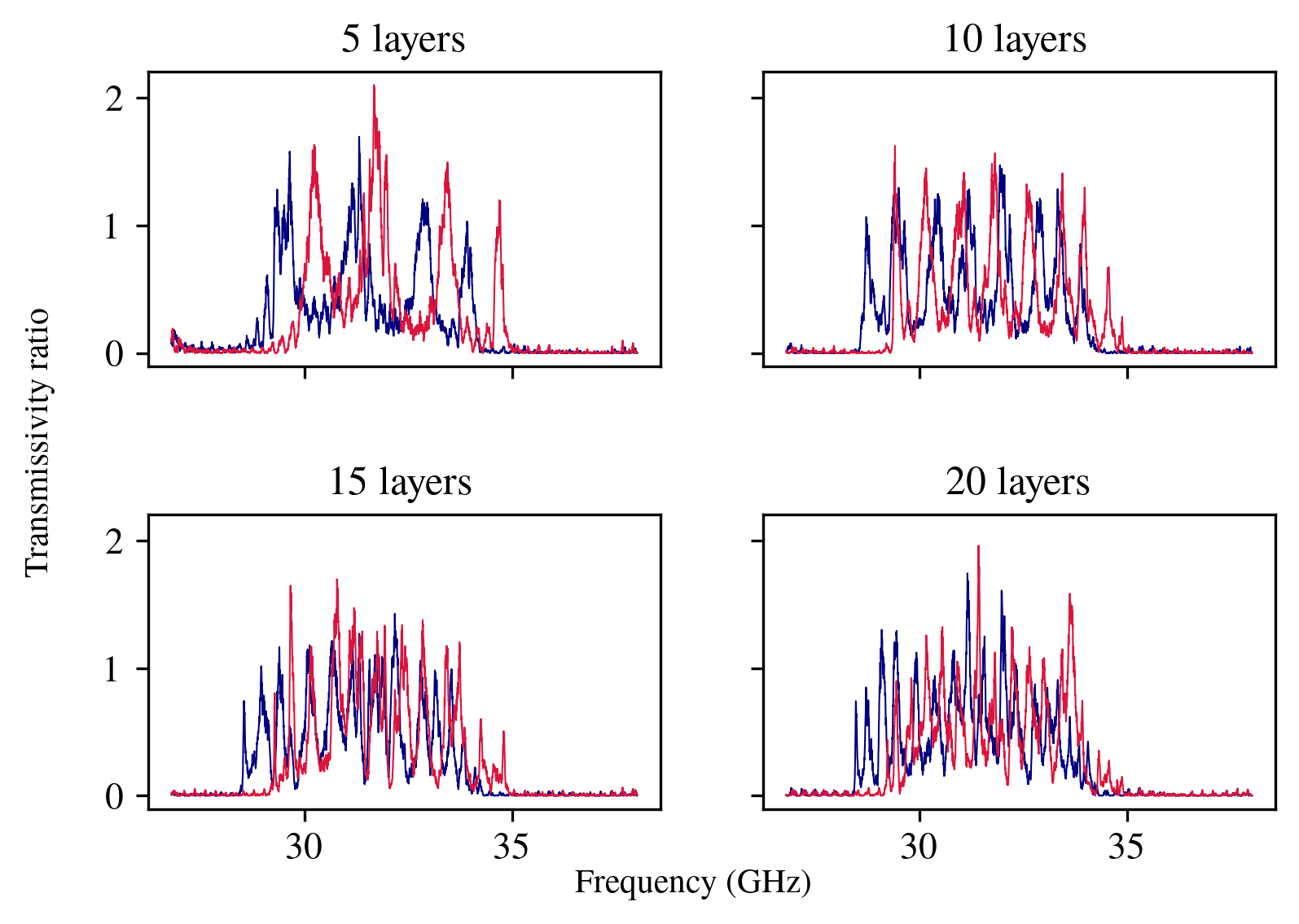}
    \caption{Top: The result of combining two free-space VNA measurements (background and two-layer Fabry-Pérot) is reported as a transmissivity magnitude ratio for a plate spacing of 6.21 mm (blue) and 6.04 mm (red). The dark curves represent transmissivities after Tukey time-domain gating to eliminate reflections and light curves represent raw data. Bottom: The measurement combination process is repeated for additional layers and the result is presented as raw data for a layer spacing of 6.21 mm (blue) and 6.04 mm (red). Time-domain gating has not been applied to prevent obscuration of small features in the resonance patterns for a high number of layers.}
    \label{fig_5}
\end{figure}

Broadband two-layer transmission measurements in both holders have consistently shown transmissivity patterns of a FP resonator with resonant frequencies 475 MHz apart as expected from their dissimilar spacings. Peak transmissivities are close to unity, as represented in Fig. \ref{fig_5}, indicating near-transparency of the resonator at its resonant frequency.

   It has been consistently observed that as the number of layers $N$ is increased, $N - 1$ roughly evenly spaced and increasingly narrow resonant structures are present in the transmissivity plots, which are distributed in a transparency band spanning approximately 6 GHz, as shown in Fig. \ref{fig_5}.

\begin{figure}[h]
    \centering
    \includegraphics[width=0.85\textwidth]{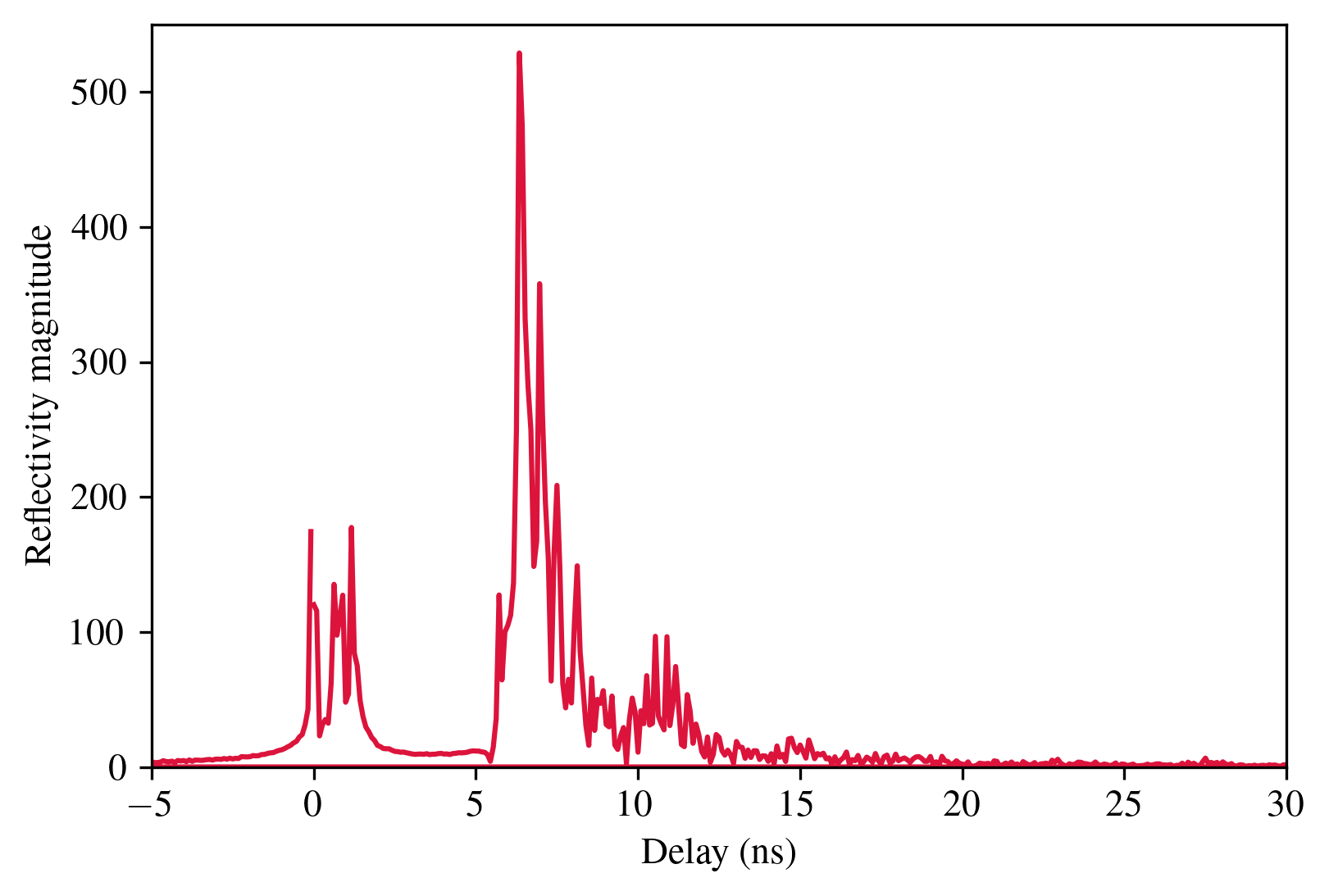}
    \caption{Time-domain reflectivity measurements for a four-layer resonator with top mirror.}
    \label{fig_6}
\end{figure}

\begin{figure}[h]
    \centering
    \includegraphics[width=0.85\textwidth]{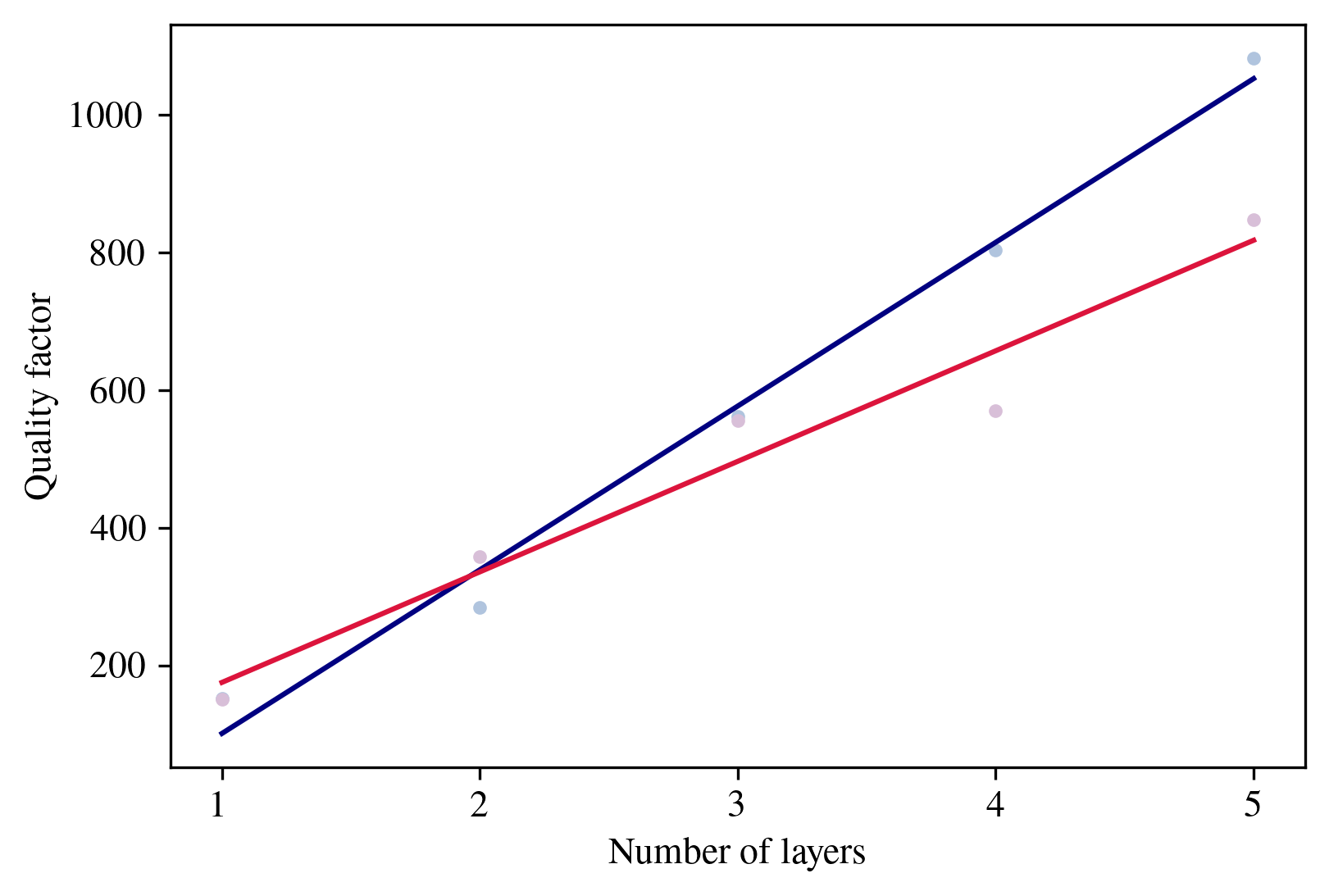}
    \caption{Quality factor calculated as $Q = \omega \tau_g$ for layer counts from one to five and for a layer spacing of 6.21 mm (blue) and 6.04 mm (red). The slope of the fit line is 238 per layer (6.21 mm) and 161 per layer (6.04 mm).}
    \label{fig_7}
\end{figure}

 As shown in Fig. \ref{fig_6}, artifacts are present in reflectivity measurements: antenna waveguide reflections—structure present around 0 ns delay—and back-reflections—signals at delays greater than $\sim$10 ns delay. Resonance is apparent in the time domain as a decaying exponential curve located at $\sim$6 ns, whose decay rate becomes slower with greater resonance originated by an increased number of plates, impeding the accurate isolation of the resonant structure for $N \gtrsim 5$ as a result \cite{MADMAX:2020}. Consequently,  measurements with one to five layers have been used for the linear fit in Fig. \ref{fig_7}. 

\section{Summary and discussion}
\label{sec:4}
Axion as a strong candidate for Galactic dark matter. DALI stands as a promising proposal to make this pseudoscalar visible. In this work, we have tested two fixed-plate optical demonstrators of the haloscope, examining their resonant behaviour. 

The persistent relation between the number of layers of the resonator and the number of resonant peaks present in the transmissivity ratio graphs seen in Fig. \ref{fig_5} suggests that resonance is present even for a high number of layers, even though its direct observation is not possible by means of the setup used in this work given that the resonant structure cannot be isolated from back-reflections for $N \gtrsim 5$. In future work we will investigate different strategies to overcome this technical limitations. The results shown in see Fig. \ref{fig_7} indicate that the ratio between the quality factor and the number of layers scales, roughly, linearly, as the models predict—$R^2 = 0.99$ (6.21 mm) and $R^2 = 0.95$ (6.04 mm)—, which could allow us to extrapolate the measurements from the recoverable resonant structures in the time domain to a larger number of layers. This fit yields a quality factor of 4623 (6.21 mm plate spacing; 3801 to 5445 at 90\% confidence interval) and 3228 (6.04 mm; 2097 to 4360 at 90\% confidence interval) for 20 layers.

\begin{figure}[h]
    \centering
    \includegraphics[width=0.85\textwidth]{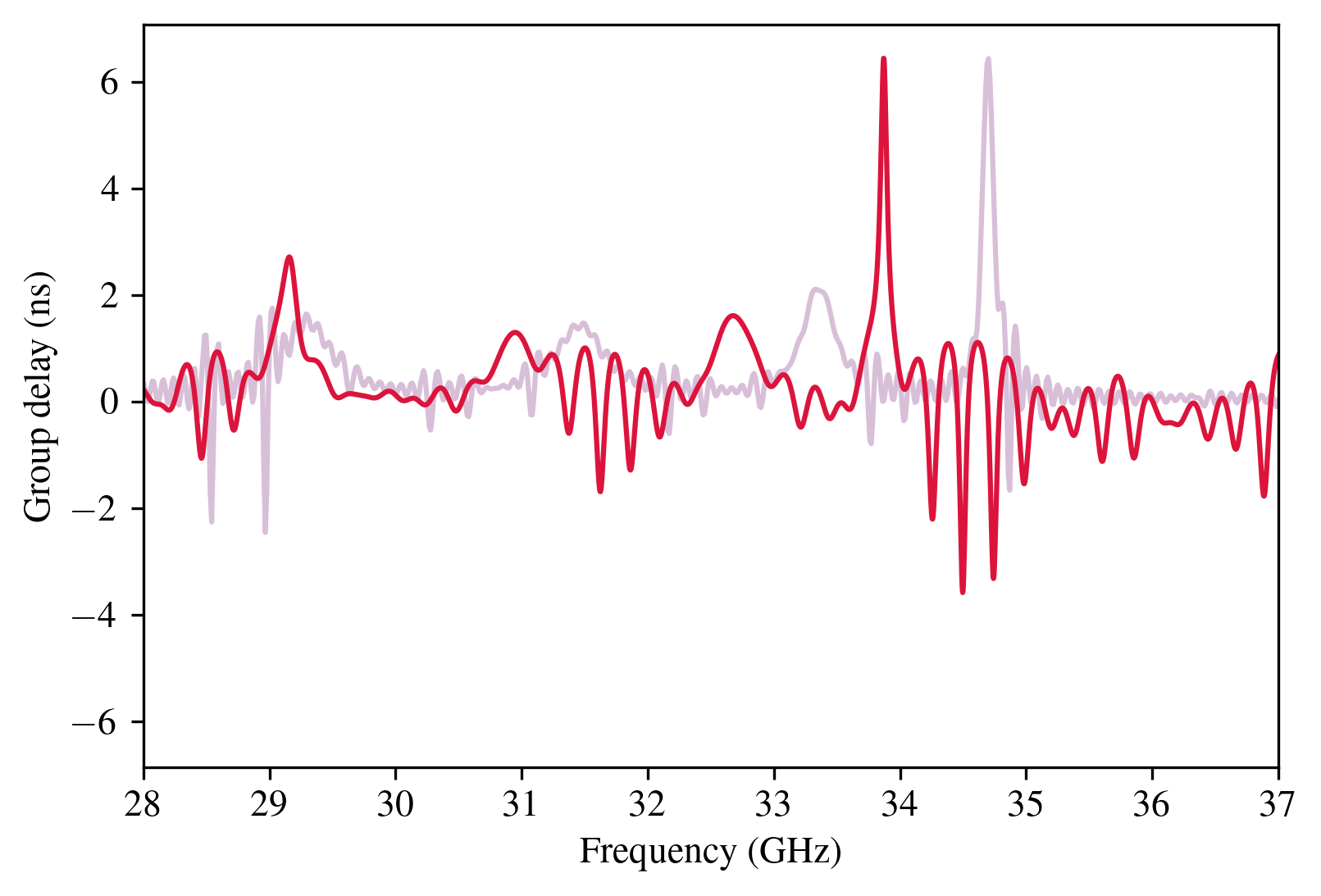}
    \caption{Measured (dark red) and simulated (light red) $S_{11}$ group delay for a four-layer Fabry-Pérot resonator. }
    \label{fig_8}
\end{figure}

The four-layer reflectivity measurement has been confronted with an one-port FEM simulation using CST Studio Suite® including the top mirror at 10 cells per wavelength. The FP resonator has been modeled including the four zirconia plates, a top mirror and a plane-wave port. The $S_{11}$ group delay has been calculated by post-processing the simulation data and plotted against the measured group delay in Fig. \ref{fig_8}. It can be observed that the measured and simulated group delay features are nearly identical in height and shape. There is a frequency shift of $\sim$800 MHz whose origin is uncertain. A similar frequency difference between measurements and FEM simulations has been observed before \cite{DeMiguel:2021dmk}. Once this frequency shift is corrected, FEM simulations can emerge as a powerful design tool.

Together, the results suggests that a quality factor of the order of $Q \sim 10^4$ is tenable over a sweep bandwidth of several dozen megahertz.

\acknowledgments
J.F.H.C. is supported by the Resident Astrophysicist Programme of the Instituto de Astrofísica de Canarias (IAC). The work of J.D.M. was supported by RIKEN's program for Special Postdoctoral Researchers (SPDR)—Project Code: 202101061013. We gratefully acknowledge financial support from the Severo Ochoa Program for Technological Projects and Major Surveys 2020-2023 under Grant No. CEX2019-000920-S; Recovery, Transformation and Resiliency Plan of Spanish Government under Grant No. C17.I02.CIENCIA.P5; FEDER operational programme under Grant No. EQC2019-006548-P; IAC Plan de Actuación 2022. We thank F. Gracia and R. Hoyland for invaluable help.



\begin{thebibliography}{99}

\bibitem{PhysRevLett.40.223}
S.~Weinberg, \emph{A new
  light boson?}, Phys. Rev. Lett. 40 (1978) 223--226.
  \href {http://dx.doi.org/10.1103/PhysRevLett.40.223}
  {\path{doi:10.1103/PhysRevLett.40.223}}.

\bibitem{PhysRevLett.40.279}
F.~Wilczek, \emph{Problem
  of strong $p$ and $t$ invariance in the presence of instantons}, Phys. Rev.
  Lett. 40 (1978) 279--282.
  \href {http://dx.doi.org/10.1103/PhysRevLett.40.279}
  {\path{doi:10.1103/PhysRevLett.40.279}}.

\bibitem{PhysRevLett.38.1440}
R.~D. Peccei, H.~R. Quinn,
  \emph{$\mathrm{CP}$
  conservation in the presence of pseudoparticles}, Phys. Rev. Lett. 38 (1977)
  1440--1443.
  \href {http://dx.doi.org/10.1103/PhysRevLett.38.1440}
  {\path{doi:10.1103/PhysRevLett.38.1440}}.

\bibitem{ABBOTT1983133}
L.~Abbott, P.~Sikivie,
  \emph{A
  cosmological bound on the invisible axion}, Physics Letters B 120~(1) (1983)
  133 -- 136.
  \href
  {http://dx.doi.org/https://doi.org/10.1016/0370-2693(83)90638-X}
  {\path{doi:https://doi.org/10.1016/0370-2693(83)90638-X}}.

\bibitem{DINE1983137}
M.~Dine, W.~Fischler,
  \emph{The
  not-so-harmless axion}, Physics Letters B 120~(1) (1983) 137 -- 141.
  \href
  {http://dx.doi.org/https://doi.org/10.1016/0370-2693(83)90639-1}
  {\path{doi:https://doi.org/10.1016/0370-2693(83)90639-1}}.

\bibitem{PRESKILL1983127}
J.~Preskill, M.~B. Wise, F.~Wilczek,
  \emph{Cosmology
  of the invisible axion}, Physics Letters B 120~(1) (1983) 127 -- 132.
  \href
  {http://dx.doi.org/https://doi.org/10.1016/0370-2693(83)90637-8}
  {\path{doi:https://doi.org/10.1016/0370-2693(83)90637-8}}.

\bibitem{Baker:2013zta}
K.~Baker, G.~Cantatore, S.~A.~Cetin, M.~Davenport, K.~Desch, B.~D\"obrich, H.~Gies, I.~G.~Irastorza, J.~Jaeckel and A.~Lindner, \textit{et al.}
Annalen Phys. \textbf{525} (2013), A93-A99
doi:10.1002/andp.201300727
[arXiv:1306.2841 [hep-ph]].

\bibitem{Irastorza:2018dyq}
I.~G.~Irastorza and J.~Redondo,
Prog. Part. Nucl. Phys. \textbf{102} (2018), 89-159
doi:10.1016/j.ppnp.2018.05.003
[arXiv:1801.08127 [hep-ph]].

\bibitem{Ciaran}
C. O'Hare,
  \emph{Axionlimits}, Zenodo, July 7, 2020.
  \href {https://github.com/cajohare/AxionLimits#readme}
  {\path{doi:10.5281/zenodo.3932430}}.

\bibitem{Sikivie:1983ip}
P.~Sikivie,
Phys. Rev. Lett. \textbf{51}, 1415-1417 (1983)
[erratum: Phys. Rev. Lett. \textbf{52}, 695 (1984)]
doi:10.1103/PhysRevLett.51.1415

\bibitem{DeMiguel:2020rpn}
J.~De Miguel,
``A dark matter telescope probing the 6 to 60 GHz band,''
JCAP \textbf{04}, 075 (2021)
doi:10.1088/1475-7516/2021/04/075
[arXiv:2003.06874 [physics.ins-det]].

\bibitem{DeMiguel:2023nmz}
J.~De Miguel and J.~F.~Hern\'andez-Cabrera,
``Discovery prospects with the Dark-photons \& Axion-Like particles Interferometer—part I,''
[arXiv:2303.03997 [hep-ph]].



\bibitem{Renk:2012}
K.~F.~Renk, Basics of Laser Physics. Springer Berlin Heidelberg, 2012. doi: 10.1007/978-3-642-23565-8


\bibitem{Okun}
L.~Okun, \emph{The limits of electrodynamics - paraphotons}, Zhurnal Eksperimental noi
  i Teoreticheskoi Fiziki 83 (1982) 892--898.

\bibitem{Vilenkin}
A.~Vilenkin, \emph{Particles and the universe}, North Holland. G. Lazarides and Q.
  Shafi (eds.) Amsterdam (1986) p. 133.
  


\bibitem{MADMAX:2019pub}
P.~Brun \textit{et al.} [MADMAX],
Eur. Phys. J. C \textbf{79} (2019) no.3, 186
doi:10.1140/epjc/s10052-019-6683-x
[arXiv:1901.07401 [physics.ins-det]].

\bibitem{Cervantes:2022epl}
R.~Cervantes, G.~Carosi, C.~Hanretty, S.~Kimes, B.~H.~LaRoque, G.~Leum, P.~Mohapatra, N.~S.~Oblath, R.~Ottens and Y.~Park, \textit{et al.}
Phys. Rev. D \textbf{106} (2022) no.10, 102002
doi:10.1103/PhysRevD.106.102002
[arXiv:2204.09475 [hep-ex]].


\bibitem{Chiles:2021gxk}
J.~Chiles, I.~Charaev, R.~Lasenby, M.~Baryakhtar, J.~Huang, A.~Roshko, G.~Burton, M.~Colangelo, K.~Van Tilburg and A.~Arvanitaki, \textit{et al.}
Phys. Rev. Lett. \textbf{128} (2022) no.23, 231802
doi:10.1103/PhysRevLett.128.231802
[arXiv:2110.01582 [hep-ex]].


\bibitem{Primakoff:1951iae}
H.~Primakoff,
Phys. Rev. \textbf{81} (1951), 899
doi:10.1103/PhysRev.81.899


\bibitem{Millar}
A.~Millar, Theoretical foundations of dielectric haloscopes. Ludwig-Maximilians-Universit{\"a}t M{\"u}nchen, 2018. http://nbn-resolving.de/urn:nbn:de:bvb:19-223692


\bibitem[De Miguel-Hern{\'a}ndez and Hoyland(2019)]{2019JInst..14R8001D} De Miguel-Hern{\'a}ndez, J., Hoyland, R.~J.\ 2019.\ Fundamentals of horn antennas with low cross-polarization levels for radioastronomy and satellite communications.\ Journal of Instrumentation 14, R08001. doi:10.1088/1748-0221/14/08/R08001


\bibitem{Molla}
Mollá, J. and Heidinger, R. and Ibarra, A. and Link, G., Dielectric properties of alumina/zirconia composites at millimeter wavelengths. Journal of Applied Physics, 73, 11, 7667-7671, 1993/06. https://doi.org/10.1063/1.353963

\bibitem{QUIJOTE:2015npn}
R.~Genova-Santos \textit{et al.} [QUIJOTE],
[arXiv:1504.03514 [astro-ph.CO]].



\bibitem{Prabhu:2018}
K.~M.~M.~Prabhu, Window Functions and Their Applications in Signal Processing, 2018. doi: 10.1201/9781315216386

\bibitem{MADMAX:2020}
J.~Egge \textit{et al.},
``A First Proof Of Principle Booster Setup For The MADMAX Dielectric Haloscope,''
[arXiv:2001.04363v2 [physics.ins-det]].

\bibitem{DeMiguel:2021dmk}
J.~De Miguel, C.~Franceschet, S.~Realini and P.~Fuerte-Rodr\'\i{}guez,
JINST \textbf{17} (2022) no.06, P06041
doi:10.1088/1748-0221/17/06/P06041
[arXiv:2108.05648 [physics.ins-det]].

  
\end{thebibliography}
\end{document}